\documentclass[11pt,epsf,letterpaper]{article}%
\usepackage{color}
\usepackage{amsmath}
\usepackage{amsfonts}
\usepackage{verbatim}
\usepackage{amssymb}
\usepackage{graphicx}
\usepackage{epstopdf}
\usepackage{mathrsfs}
\usepackage{epsfig}
\usepackage{cite}
\providecommand{\U}[1]{\protect\rule{.1in}{.1in}}
\newcommand{\bl}{\boldsymbol}

\newcommand{\ph}{\phantom}

\begin{document}

\title{On Spaces with the Maximal Number of Conformal Killing Vectors}
\author{Carlos Batista\\\textit{Departamento de F\'\i sica, Universidade Federal de
Pernambuco,} \\\textit{50740-560 Recife-PE, Brazil}    \\ carlosbatistas@df.ufpe.br }
\maketitle

%\title{An Exact Solution for Einstein-Maxwell Equation in Six Dimensions}
%\author{Andr\'{e}s Anabal\'{o}n$^{(1)}$ and Carlos Batista$^{(2)}$\\\textit{$^{(1)}$Departamento de Ciencias, Facultad de Artes Liberales y}\\\textit{Facultad %de Ingenier\'{\i}a y Ciencias, Universidad Adolfo
%Ib\'{a}\~{n}ez,}\\\textit{Av. Padre Hurtado 750, Vi\~{n}a del Mar, Chile}\\\textit{$^{(2)}$Departamento de F\'\i sica, Universidade Federal de
%Pernambuco,} \\\textit{50670-901 Recife-PE, Brazil} \\andres.anabalon@uai.cl, carlosbatistas@df.ufpe.br}
%\maketitle

\begin{abstract}
It is natural to expect and simple to prove that every conformally flat space possess the maximal number of conformal Killing vector fields (CKVs). On the other hand, it is interesting to ask whether the converse is true. Is conformal flatness a necessary condition for the existence of the maximal number of CKVs? In this review article it is proven that the answer is yes, a space admits the maximal number of CKVs if, and only if, it is conformally flat.
\end{abstract}

\section{Introduction}\label{Sec.Introduc}

Just as the Killing vector fields are generators of isometries, diffeomorphisms that preserve the metric, the conformal Killing vector fields (CKVs) are the generators of conformal transformations, namely diffeomorphisms that preserve the metric apart from a multiplicative scale factor. Due to the central role played by conformal field theories in modern theoretical physics, the interest in CKVs is increasing. Indeed, conformal field theories are on the basis of the study of critical phenomena and the associated renormalization group \cite{Ginsparg:1988ui,DiFrancesco}. Moreover, conformal field theories comprise an essential tool for the AdS/CFT correspondence \cite{Maldacena:1997re}.

From the geometrical point of view, CKVs lead to first integrals along null geodesics, which it is of great utility for the integration of the null geodesic equation.  In addition, CKVs can be used to construct Killing tensors of order two \cite{Koutras92}, which lead to conserved charges quadratic in the momentum along any geodesic, even non-null ones. However, it is worth pointing out that not all Killing tensors of order two can be constructed out of a CKV, as exemplified by the nontrivial Killing tensor of the Kerr metric \cite{Jerie99}. Moreover, Killing-Yano tensors of order $n-1$ can also be constructed out of special types CKVs \cite{KYn-1}. Thus, CKVs are of great relevance to attain the integration of geodesic equation.

It is well-known that a space of dimension $n$ can have up to $\frac{1}{2}n(n+1)$ independent Killing vectors and that this maximal number of CKVs is attained just for spaces of constant curvature \cite{Weinberg-Book}. These spaces are, thus, dubbed maximally symmetric. For instance, the maximally symmetric spaces of Lorentzian signature are the Minkowski space (zero curvature), the de Sitter space (positive curvature) and the anti-de Sitter space (negative curvature). In an analogous fashion, it would be interesting to characterize the spaces possessing the maximal number of CKVs. This is the aim of this review article. As we shall see in the sequel, for dimension $n\geq3$, the maximal number of CKVs is $\frac{1}{2}(n+1)(n+2)$, while for $n=2$ any space will admit infinity independent CKVs. The main goal of this paper is to find the necessary and sufficient conditions for a curved space, of arbitrary dimension and signature, to admit the maximal number of independent CKVs. More precisely, it will be proved that this maximal number is attained if, and only if, the space is conformally flat.
This article is being referred to as a review article because a proof of the latter assertion is already available in the literature, in an appendix of a 1949 book of Eisenhart \cite{EisenhartBook}. However, besides the latter reference, the present author is not aware of any other proof of this theorem. Even the simple statement of the theorem is hardly found in the books of general relativity, one exception being Stephani's et. al book \cite{Stephani}. The difficulty in finding a proof of such theorem was the motivation for writing this paper, in which we try to follow a didactic and self-contained presentation. Here, we provide a different and, hopefully, clearer proof of the mentioned theorem.  In addition, we provide  insights on the interpretations of the different degrees of freedom of a CKV.

The outline of the article is the following. In Sec. \ref{Sec.BriefRev} we recall the definition of a CKV and provide a brief review of its basic properties. Then, in Sec. \ref{Sec.FlatSpaces}, we fully and explicitly integrate the CKV equation in flat spaces of arbitrary dimension and signature. The latter integration helps to build the intuition about the degrees of freedom behind a CKV. In particular we present the interpretation of the several parts that compose the general CKV. Finally, in Sec. \ref{Sec.CurvedSpaces} we show that the maximal number of CKVs in a space of dimension $n$ is $\frac{1}{2}(n+1)(n+2)$  and prove that a space admit this maximal number of CKVs if, and only if, it is conformally flat.

\section{Brief Review of CKVs}\label{Sec.BriefRev}

In this section we shall define the conformal Killing vector fields (CKVs) and workout some of its main properties. But, before we head to the main content, let us establish the conventions and notation used throughout the paper. In this article we are assuming that our manifold has dimension $n$, with $n\geq 2$, and that it is endowed with a metric $\bl{g}$ of arbitrary signature. Moreover, we shall use the Levi-Civita connection, here denoted by $\bl{\nabla}$. Therefore, henceforth, whenever we write ``a space'' the reader should understand as a manifold endowed with metric and the Levi-Civita connection. As usual, indices enclosed by round brackets are assumed to be symmetrized, while indices enclosed by square brackets means that they are antisymetrized.

A vector field $\bl{\xi}$ is said to be a conformal Killing vector (CKV) when it obeys the following differential equation
\begin{equation}\label{CKVeq0}
  \nabla_{(a} \xi_{b)} =  \frac{1}{2}( \nabla_a \xi_b +   \nabla_b \xi_a) = f\,g_{ab} \,,
\end{equation}
where $f$ is some scalar function. An important applicability for these objects is that they generate conserved charges along null geodesics. Indeed, if $\bl{N}$ is a vector field tangent to a congruence of affinely parameterized null geodesics, $N^a N_a=0$ and $N^a\nabla_a N^b=0$, then, whenever $\bl{\xi}$ is a CKV, the scalar $N^a\xi_a$ is a constant of motion along these geodesics, i.e., $N^b\nabla_b(N^a\xi_a)=0$. This conservation law provides a first order differential equation for the parameterized geodesic, helping on the integration of the geodesic equation, which is of second order.

Another important characterization of the CKVs is that they are the generators of the conformal symmetries of the metric. Indeed, suppose we perform a coordinate transformation representing an infinitesimal translation along the vector field $\bl{\xi}$, namely
\begin{equation}\label{Transf.x}
  x^{\alpha} \mapsto \tilde{x}^{\alpha} = x^{\alpha} + \epsilon \,\xi^\alpha \,,
\end{equation}
where $\{x^\alpha\}$ is the original coordinate system while $\{\tilde{x}^\alpha\}$ is the new coordinate system, with $\epsilon$ being a constant infinitesimal parameter. In the new coordinates the components of the metric are given by
\begin{equation*}
  \tilde{g}_{\alpha\beta}(x +  \epsilon \,\xi) =  \tilde{g}_{\alpha\beta}(\tilde{x}) = g_{\sigma\rho}(x) \,
  \frac{\partial x^\sigma}{\partial \tilde{x}^\alpha} \frac{\partial x^\rho}{\partial \tilde{x}^\beta} =
  g_{\alpha\beta}(x) -\epsilon \,\left[ g_{\alpha\sigma}(x) \partial_\beta \xi^\sigma +  g_{\beta\sigma}(x) \partial_\alpha \xi^\sigma \right] + O(\epsilon^2) \,.
\end{equation*}
Now, expanding the left hand side of the latter equation in Taylor series, we end up with the following relation
\begin{equation}\label{Transf.g}
  \tilde{g}_{\alpha\beta}(x) =  g_{\alpha\beta}(x) - 2 \epsilon \, \nabla_{(\alpha} \xi_{\beta)} + O(\epsilon^2)  \,.
\end{equation}
Thus, up to first order in $\epsilon$, the functional forms of $\tilde{g}_{\alpha\beta}$ and  $g_{\alpha\beta}$ will be related to a conformal transformation ($\tilde{g}_{\alpha\beta} \propto g_{\alpha\beta}$ ) if, and only if, $\bl{\xi}$ obeys the CKV equation.

Since the CKVs are the generators of the conformal symmetries, it is natural to expect that spaces that are conformally related have a correspondence between their CKVs.  In order to check this, note that when we perform a conformal transformation
\begin{equation*}
  g_{ab} \,\mapsto\, \tilde{g}_{ab} = e^{2\omega} g_{ab} \,,
\end{equation*}
where $\omega$ is some function on the space, the components of the Christoffel symbol get changed as follows
\begin{equation*}
  \Gamma^{c}_{ab} \,\mapsto\,  \tilde{\Gamma}^{c}_{ab} =  \Gamma^{c}_{ab} +  \delta\Gamma^{c}_{ab}  \;,\; \textrm{ where }\;
  \delta\Gamma^{c}_{ab} = \left( \delta^c_a \,\partial_b\omega + \delta^c_b \,\partial_a\omega - g_{ab}  g^{ch}\partial_h\omega   \right)\,.
\end{equation*}
Now, let $\xi^a$ be a CKV on the space with metric $g_{ab}$. Then, it follows that $\tilde{\xi}^a = \xi^a$ is a CKV on the space with metric $\tilde{g}_{ab}$. Indeed, since $\tilde{\xi}_b = \tilde{g}_{bh} \tilde{\xi}^h = e^{2\omega} \xi_b$, it follows that
\begin{equation*}
  \tilde{\nabla}_a \tilde{\xi}_b = \nabla_a \tilde{\xi}_b - \delta\Gamma^{c}_{ab} \tilde{\xi}_c =
  \nabla_a (e^{2\omega} \xi_b) - e^{2\omega} \delta\Gamma^{c}_{ab} \xi_c = e^{2\omega} \left( \nabla_a \xi_b + \xi_{[b}\partial_{a]}\omega +
  g_{ab} \xi^c\partial_c\omega  \right)\,.
\end{equation*}
Thus, taking the symmetric part of the latter equation and using Eq. \eqref{CKVeq0}, we find
\begin{equation*}
  \tilde{\nabla}_{(a} \tilde{\xi}_{b)} = e^{2\omega} \left( f +  \xi^c\partial_c\omega  \right) g_{ab}  =
  \tilde{f}\, \tilde{g}_{ab}   \;,\; \textrm{ where }\;   \tilde{f} = \left( f +  \xi^c\partial_c\omega  \right)\,,
\end{equation*}
proving that CKVs are conformally invariant.

%%%%%%%%%%%%%%%%%%%%%%%%%%%%%%%%%%%%%%%%%%%%%%%%%%%%%%%%%%%%%%%%%%%%%%%%%%%%%%%%%%%%%%%%%%%%%%%%
%%%%%%%%%%%%%%%%%%%%%%%%%%%%%%%%%%%%%%%%%%%%%%%%%%%%%%%%%%%%%%%%%%%%%%%%%%%%%%%%%%%%%%%%%%%%%%%%
%%%%%%%%%%%%%%%%%%%%%%%%%%%%%%%%%%%%%%%%%%%%%%%%%%%%%%%%%%%%%%%%%%%%%%%%%%%%%%%%%%%%%%%%%%%%%%%%
%%%%%%%%%%%%%%%%%%%%%%%%%%%%%%%%%%%%%%%%%%%%%%%%%%%%%%%%%%%%%%%%%%%%%%%%%%%%%%%%%%%%%%%%%%%%%%%%
%%%%%%%%%%%%%%%%%%%%%%%%%%%%%%%%%%%%%%%%%%%%%%%%%%%%%%%%%%%%%%%%%%%%%%%%%%%%%%%%%%%%%%%%%%%%%%%%
%%%%%%%%%%%%%%%%%%%%%%%%%%%%%%%%%%%%%%%%%%%%%%%%%%%%%%%%%%%%%%%%%%%%%%%%%%%%%%%%%%%%%%%%%%%%%%%%
%%%%%%%%%%%%%%%%%%%%%%%%%%%%%%%%%%%%%%%%%%%%%%%%%%%%%%%%%%%%%%%%%%%%%%%%%%%%%%%%%%%%%%%%%%%%%%%%
%%%%%%%%%%%%%%%%%%%%%%%%%%%%%%%%%%%%%%%%%%%%%%%%%%%%%%%%%%%%%%%%%%%%%%%%%%%%%%%%%%%%%%%%%%%%%%%%
%%%%%%%%%%%%%%%%%%%%%%%%%%%%%%%%%%%%%%%%%%%%%%%%%%%%%%%%%%%%%%%%%%%%%%%%%%%%%%%%%%%%%%%%%%%%%%%%
%%%%%%%%%%%%%%%%%%%%%%%%%%%%%%%%%%%%%%%%%%%%%%%%%%%%%%%%%%%%%%%%%%%%%%%%%%%%%%%%%%%%%%%%%%%%%%%%

\section{CKVs of a Flat Space}\label{Sec.FlatSpaces}

In order to gain some intuition about the general case of curved spaces, it is instructive to first deal with the conformal Killing vectors (CKVs) of flat spaces. In this section we will explicitly integrate the CKV equation in a flat space of arbitrary dimension and arbitrary signature.

In a space of zero curvature, a flat space, we can always introduce coordinates $\{x^\alpha\}$ such that the components of the metric in the coordinate frame are constant,
$$ \bl{g}(\partial_\alpha, \partial_\beta) = \eta_{\alpha \beta} \;,\;\; \partial_\sigma \eta_{\alpha \beta} = 0.   $$
In these coordinates, the components of the Christoffel symbol are all vanishing, so that the covariant derivative reduces to the partial derivative. In the remainder of the section, we shall adopt this kind of coordinate system.

%For instance, if the coordinates are chosen in a way that the matric $\eta_{\alpha\beta}$ is diagonal and with $\pm1$ in the diagonal, we say that the %coordinates $\{x^\alpha\}$ are cartesian.

Assuming that $\xi_a$ is a smooth field, it can be expanded in Taylor series around the origin of the coordinate system,
\begin{equation}\label{XiTaylor}
  \xi_\alpha = A_\alpha + A_{\alpha,\beta_1} x^{\beta_1} + \frac{1}{2} A_{\alpha,\beta_1\beta_2} x^{\beta_1} x^{\beta_2} +
  \frac{1}{3}  A_{\alpha,\beta_1\beta_2\beta_3} x^{\beta_1} x^{\beta_2} x^{\beta_3} + \cdots \,,
\end{equation}
where $A_{\alpha,\beta_1\cdots \beta_p}$ are constants that are defined by the derivatives of the field $\xi_a$ at the origin. Note that we can consider that these constants are totally symmetric on the indices $\beta_1\cdots \beta_p$,
\begin{equation}\label{ASym}
  A_{\alpha,\beta_1\cdots \beta_p} = A_{\alpha,(\beta_1\cdots \beta_p)} \,,
\end{equation}
inasmuch as any skew-symmetric part would not contribute to the expansion \eqref{XiTaylor}, since $x^{\beta_1} \cdots x^{\beta_p}$ is totally symmetric. Then, taking the covariant derivative of $\bl{\xi}$, we find
\begin{equation}\label{DXiTaylor}
  \nabla_\sigma \xi_\alpha = \partial_\sigma \xi_\alpha
  = A_{\alpha,\sigma} +  A_{\alpha,\sigma \beta_2} x^{\beta_2} +  A_{\alpha,\sigma \beta_2 \beta_3} x^{\beta_2} x^{\beta_3} + \cdots \,.
\end{equation}

Now, let us impose the CKV equation for the vector field $\bl{\xi}$, namely we shall integrate the equation $\nabla_{(\sigma}\xi_{\alpha)} = f \eta_{\sigma \alpha}$. In order to implement this, we shall expand the function $f$ in Taylor series,
\begin{equation}\label{fTaylor}
  f = B + B_{\beta_1} x^{\beta_1} +  B_{\beta_1 \beta_2} x^{\beta_1} x^{\beta_2} +
  B_{\beta_1 \beta_2 \beta_3} x^{\beta_1} x^{\beta_2} x^{\beta_3} + \cdots \,,
\end{equation}
where the coefficients $B$ and $B_{\beta_1\cdots \beta_p}$ are constants with the latter being totally symmetric on the indices $\beta_1\cdots \beta_p$,
\begin{equation}\label{BSym}
  B_{\beta_1\cdots \beta_p} = B_{(\beta_1\cdots \beta_p)} \,.
\end{equation}
Then, using Eqs. \eqref{DXiTaylor} and \eqref{fTaylor} and comparing both sides of the equation $\nabla_{(\sigma}\xi_{\alpha)} = f \eta_{\sigma \alpha}$ order to order in the power series, lead us to the following constraints:
\begin{equation}\label{CostraintFlat}
  A_{(\alpha,\sigma)} = B \eta_{\alpha\sigma} \quad \textrm{and} \quad
  A_{(\alpha,\sigma)\beta_2\cdots \beta_p} = B_{\beta_2\cdots \beta_p} \eta_{\alpha\sigma}\,.
\end{equation}
Note that no restriction is imposed over $A_\alpha$, so that it can be arbitrary. Since any tensor with two indices can be decomposed as the sum of its symmetric part and its anti-symmetric part, it follows that the general solution for the constraint (\ref{CostraintFlat}) over $A_{\alpha,\sigma}$ is
\begin{equation}\label{A2}
  A_{\alpha,\beta} =  B\, \eta_{\alpha\beta} + C_{\alpha\beta} \,,
\end{equation}
where $C_{\alpha\beta} =  C_{[\alpha\beta]}$ is an arbitrary skew-symmetric tensor with constant components in this coordinate frame. Concerning the components $A_{\alpha,\sigma\beta_2\cdots \beta_p}$, the constraint (\ref{CostraintFlat}) implies that
\begin{equation}\label{Askew}
  A_{\alpha,\sigma\beta_2 \cdots \beta_p} = - A_{\sigma,\alpha \beta_2 \cdots \beta_p} + 2 \eta_{\alpha\sigma} B_{\beta_2\cdots \beta_p} \,.
\end{equation}
Thus, using this relation along with Eq. (\ref{ASym}) successive times, we find that
\begin{align*}
 A_{\alpha,\sigma\beta_2 \beta_3\cdots \beta_p} &= A_{\alpha,\beta_2 \sigma \beta_3\cdots \beta_p} \\
&=  - A_{\beta_2,\alpha \sigma \beta_3\cdots \beta_p} + 2 \eta_{\alpha\beta_2} B_{\sigma \beta_3\cdots \beta_p} \\
 &= - A_{\beta_2,\sigma \alpha \beta_3\cdots \beta_p} + 2 \eta_{\alpha\beta_2} B_{\sigma \beta_3\cdots \beta_p} \\
 &=  A_{\sigma,\beta_2 \alpha \beta_3\cdots \beta_p} - 2 \eta_{\sigma\beta_2} B_{\alpha \beta_3\cdots \beta_p} + 2 \eta_{\alpha\beta_2} B_{\sigma \beta_3\cdots \beta_p}  \\
 &=  A_{\sigma, \alpha \beta_2 \beta_3\cdots \beta_p} - 2 \eta_{\sigma\beta_2} B_{\alpha \beta_3\cdots \beta_p} + 2 \eta_{\alpha\beta_2} B_{\sigma \beta_3\cdots \beta_p} \\
 &= - A_{\alpha,\sigma\beta_2 \beta_3\cdots \beta_p} + 2 \eta_{\alpha\sigma} B_{\beta_2 \beta_3\cdots \beta_p}  - 2 \eta_{\sigma\beta_2} B_{\alpha \beta_3\cdots \beta_p} + 2 \eta_{\alpha\beta_2} B_{\sigma \beta_3\cdots \beta_p} \,,
\end{align*}
which yields the following identity
\begin{equation}\label{AAgen}
  A_{\alpha,\sigma\beta_2 \beta_3\cdots \beta_p}  =   \eta_{\alpha\sigma} B_{\beta_2 \beta_3\cdots \beta_p} +  \eta_{\alpha\beta_2} B_{\sigma \beta_3\cdots \beta_p} -  \eta_{\sigma\beta_2} B_{\alpha \beta_3\cdots \beta_p} \,.
\end{equation}

For the special case $p=2$, Eq. (\ref{AAgen}) implies that
\begin{equation}\label{AA2}
  A_{\alpha,\sigma\beta_2}  =   \eta_{\alpha\sigma} B_{\beta_2 } +  \eta_{\alpha\beta_2} B_{\sigma } -  \eta_{\sigma\beta_2} B_{\alpha } \,,
\end{equation}
which is in perfect accordance with Eq. \eqref{ASym} for an arbitrary choice of $B_\sigma$, since the right hand side of (\ref{AA2}) is symmetric on the pair of indices $\sigma\beta_2$. For the remaining cases, $p\geq3$, the right hand side of Eq. (\ref{AAgen}) is not symmetric for an arbitrary choice of $B_{\sigma \beta_3\cdots \beta_p}$. Thus, let us impose this symmetry. Interchanging the indices $\beta_2$ and $\beta_3$ in Eq. (\ref{AAgen}) lead to
\begin{equation}\label{AAgen2}
   A_{\alpha,\sigma\beta_3 \beta_2 \beta_4\cdots \beta_p}  =   \eta_{\alpha\sigma} B_{\beta_3 \beta_2  \beta_4\cdots \beta_p} +  \eta_{\alpha\beta_3} B_{\sigma \beta_2  \beta_4 \cdots \beta_p} -  \eta_{\sigma\beta_3} B_{\alpha \beta_2 \beta_4 \cdots \beta_p} \,.
\end{equation}
Now, making the difference between Eqs. (\ref{AAgen}) and (\ref{AAgen2}) and then using the identities (\ref{ASym}) and (\ref{BSym}), we find that
\begin{equation*}
    \eta_{\alpha\beta_2} B_{\sigma \beta_3  \beta_4\cdots \beta_p} -  \eta_{\sigma\beta_2} B_{\alpha \beta_3 \beta_4\cdots \beta_p}  - \eta_{\alpha\beta_3} B_{\sigma \beta_2 \beta_4\cdots \beta_p} +  \eta_{\sigma\beta_3} B_{\alpha \beta_2 \beta_4\cdots \beta_p} = 0 \,.
\end{equation*}
Contracting the latter equation with $\eta^{\alpha\beta_2}$ leads to
\begin{equation}\label{Bgen}
  (n-2) B_{\sigma \beta_3 \beta_4\cdots \beta_p} = - \eta_{\sigma\beta_3} \eta^{\alpha \rho}B_{\alpha \rho \beta_4\cdots \beta_p}\,.
\end{equation}
Since the left hand side of the latter equation vanishes when $n=2$, we conclude that the two-dimensional case is special, so that we shall consider it separately at the end of this section. Thus, in what follows we assume $n\geq 3$. Contracting Eq. (\ref{Bgen}) with $\eta^{\sigma\beta_3}$ lead us to the conclusion that
$\eta^{\alpha \rho}B_{\alpha \rho \beta_4\cdots \beta_p}$ vanishes. Taking this into account in Eq. (\ref{Bgen}), yields
\begin{equation}\label{Bzero}
  B_{\beta_2 \beta_3 \beta_4\cdots \beta_p} = 0 \quad \textrm{if}  \quad p\geq 3\,, \;\textrm{ and }\;  n\neq 2 \,.
\end{equation}
Then, inserting this into Eq. (\ref{AAgen}) leads to
\begin{equation}\label{Azero}
  A_{\alpha,\sigma \beta_2 \beta_3 \beta_4\cdots \beta_p} = 0 \quad \textrm{if}  \quad p\geq 3\,, \;\textrm{ and }\;  n\neq 2 \,.
\end{equation}
Thus, for $n\neq 2$, the most general CKV in a flat space is given by
\begin{equation}\label{CKV_Flat}
  \xi_\alpha = A_\alpha + \left( C_{\alpha\beta} + B \eta_{\alpha\beta} \right) x^{\beta}
  + \frac{1}{2} \left( \eta_{\alpha\beta_1} B_{\beta_2 } +  \eta_{\alpha\beta_2} B_{\beta_1 }
   -  \eta_{\beta_1\beta_2} B_{\alpha } \right) x^{\beta_1}  x^{\beta_2} \,.
\end{equation}
The function $f$ associated to this CKV is 
\begin{equation}\label{f-flat}
f= B + B_\beta x^\beta \,.
\end{equation}
The constant coefficients $A_\alpha$, $C_{\alpha\beta} = C_{[\alpha\beta]}$, $B$ and $B_\alpha$ are completely arbitrary, comprising a total of
\begin{equation*}
  n + \frac{1}{2}n(n-1) + 1 + n = \frac{1}{2}(n+1)(n+2)
\end{equation*}
free parameters.

Now, we shall try do interpret each of the different degrees of freedom of the general CKV (\ref{CKV_Flat}). For instance, let us start trying to interpret the role of the component $A_\alpha$. With this aim, we shall set $C_{\alpha\beta}$, $B$ and $B_\alpha$ equal to zero, in which case the transformations (\ref{Transf.x}) and (\ref{Transf.g}) are given by
\begin{equation}\label{Translation}
 x^{\alpha} \mapsto \tilde{x}^{\alpha} = x^{\alpha} + \epsilon \,A^{\alpha} \quad \textrm{and} \quad 
g_{\alpha\beta} \,\mapsto\, \tilde{g}_{\alpha\beta} = g_{\alpha\beta} =  \eta_{\alpha\beta}  \,,
\end{equation}
where it has been used the fact that $\nabla_{(\alpha}\xi_{\beta)}= f g_{\alpha\beta}$ along with  Eq. (\ref{f-flat}). The coordinate transformation (\ref{Translation}) represents a global translation in the direction $A^\alpha$.  Note that if $A^\alpha$ is nonzero, no point is fixed by this transformation. Now, setting  $A_{\alpha}$, $B$ and $B_\alpha$ equal to zero, it follows from Eq. (\ref{f-flat}) that $f=0$, leading to the following transformations
\begin{equation*}
 x^{\alpha} \mapsto \tilde{x}^{\alpha} = x^{\alpha} + \epsilon \,C^{\alpha}_{\ph{\alpha}\beta} x^\beta
\quad \textrm{and} \quad g_{\alpha\beta} \,\mapsto\, \tilde{g}_{\alpha\beta} =  \eta_{\alpha\beta} \,.
\end{equation*}
So, the metric is invariant under this transformation. In addition, note that the origin, $x^\alpha=0$, is fixed by the latter transformation, while all other points are moved. Thus, the transformations associated to $C_{\alpha\beta}$ preserve the inner products and have a single fixed point, these are features of a rotation. Indeed, using cartesian coordinates of the space $\mathbb{R}^n$ we can easily see that $C_{\alpha\beta}$ generate rotations. Now, let us set $A_\alpha$, $C_{\alpha\beta}$, and $B_\alpha$ equal to zero. In this case the transformations associated to the CKV are
\begin{equation*}
 x^{\alpha} \mapsto \tilde{x}^{\alpha} = (1+ \epsilon \,B)x^{\alpha} 
\quad \textrm{and} \quad g_{\alpha\beta} \,\mapsto\, \tilde{g}_{\alpha\beta} = (1- 2\epsilon \,B)  \eta_{\alpha\beta} \,,
\end{equation*}
which is a scaling transformation in the coordinates and in the metric, formally referred to as dilatation or homothety. Finally, in order to interpret the role of $B_\alpha$, let us set $A_\alpha$, $C_{\alpha\beta}$, and $B$ equal to zero, in which case the transformations (\ref{Transf.x}) and (\ref{Transf.g}) yield
\begin{equation}\label{SpecialConf}
 x^{\alpha} \mapsto \tilde{x}^{\alpha} = x^{\alpha} +  \epsilon \left(  B_{\beta} x^\beta\,x^\alpha
   - \frac{1}{2} x^\beta x_\beta\, B^{\alpha } \right) 
\quad \textrm{and} \quad g_{\alpha\beta} \,\mapsto\, \tilde{g}_{\alpha\beta} =  (1- 2\epsilon \,B_{\sigma} x^\sigma)  \eta_{\alpha\beta} \,.
\end{equation}
The latter transformations are the so-called special conformal transformations, they are a composition of a inversion a translation and another inversion, which can be understood from the fact that the following relation holds up to first order in $\epsilon$:
\begin{equation*}
  \frac{\tilde{x}^{\alpha}}{\eta_{\beta\sigma}\tilde{x}^{\beta}\tilde{x}^{\sigma}} =
 \frac{x^{\alpha}}{\eta_{\beta\sigma} x^{\beta} x^{\sigma}} - \frac{\epsilon}{2} B^\alpha\,.
\end{equation*}
Table \ref{Tab.CKV} sums up the interpretation of the different parts of the general CKV presented in Eq. (\ref{CKV_Flat}).
\begin{table}[!htbp]
\begin{center}
\begin{tabular}{c|c|c|c}
  \hline
  \textsc{Choice of Parameters}  \,&\, $\bl{\xi}$ \,&\, $f$ \,&\, \textsc{Interpretation} \\ \hline \hline
  \small{$A_\alpha = \eta_{\alpha \hat{\alpha}},\, C_{\alpha\beta} = 0,\, B =0,\, B_\alpha = 0 $} \,&\, \small{$\partial_{\hat{\alpha}}$} & \small{$0$} & \small{Translations}   \\ \hline
  \small{$A_\alpha = 0,\, C_{\alpha\beta} = (\delta_\alpha^{\hat{\alpha}}  \delta_\beta^{\hat{\beta}} - \delta_\beta^{\hat{\alpha}} \delta_\alpha^{\hat{\beta}} ),\,
   B = 0,\, B_\alpha = 0$} \,&\,\small{ $x^{\hat{\beta}} \partial^{\hat{\alpha}} -  x^{\hat{\alpha}} \partial^{\hat{\beta}} $} \,&\, \small{$0$} \,&\, \small{Rotations} \\ \hline
  \small{$A_\alpha = 0,\, C_{\alpha\beta} =0,\, B=1,\,  B_\alpha = 0 $} \,&\, \small{$x^\alpha\partial_\alpha$} \,&\, \small{$1$} \,&\, \small{Dilatation (homothety)} \\  \hline
   \small{$ A_\alpha = 0,\, C_{\alpha\beta} = 0,\, B = 0,\, B_\alpha= \delta_\alpha^{\hat{\alpha}}$} \,&\, \small{$x^{\hat{\alpha}} x^\beta\partial_\beta - \frac{1}{2}x^\beta x_\beta \partial^{\hat{\alpha}}$} \,&\, \small{$x^{\hat{\alpha}}$} \,&\, \small{Special Conf. Transf.} \\  \hline
\end{tabular}\caption{\footnotesize{This table describes the different parts of the general CKV of the flat space shown in Eq. (\ref{CKV_Flat}). The indices with hat, $\hat{\alpha}$ and $\hat{\beta}$, are assumed to be fixed indices, although arbitrary. For instance, we can picture that $\hat{\alpha}=1$ and $\hat{\beta}=2$. In the last column we provide the interpretation of the transformations generated by these vector fields. In particular, in the last row the abbreviation stands for ``special conformal transformations''. From the fact that $f=0$ for the first two vector fields, we conclude that these CKVs are, actually, Killing vector fields.  } }\label{Tab.CKV}
\end{center}
\end{table}

Now, let us consider the two-dimensional spaces. Using cartesian coordinates $\{y^1,y^2\}$, the line element of a flat space of dimension two can always be written as follows
\begin{equation*}
  ds^2 = \epsilon_1 (dy^1)^2 + \epsilon_2 (dy^2)^2 \,,
\end{equation*}
where $\epsilon_1$ and $\epsilon_2$ are $\pm1$. These constants encode the signature of the space. Then, defining the coordinates
\begin{equation}\label{zw}
  z=\frac{1}{\sqrt{2}}(\sqrt{\epsilon_1}\, y^1 + \sqrt{-\epsilon_2} \,y^2)   \quad \textrm{and} \quad
  w=\frac{1}{\sqrt{2}}(\sqrt{\epsilon_1} \,y^1 - \sqrt{-\epsilon_2} \,y^2) \,,
\end{equation}
it turns out that the line element is conveniently written as
\begin{equation*}
  ds^2 = 2\,dz \,dw \,.
\end{equation*}
Thus, performing the coordinate transformations $(z,w)\mapsto (\tilde{z},\tilde{w})$, where $z=H(\tilde{z})$ and $w=G(\tilde{w})$, with $H$ and $G$ being arbitrary nonconstant functions of their argument, it turns out that the form of the line element changes only by a conformal factor,
$$ ds^2 = H'\, G'\,   (2\,d\tilde{z} \,d\tilde{w} ) \,, $$
where the primes denote derivatives. Thus, we achieved a conformal transformation. Since, apart from being nonconstant, the functions $H$ and $G$ are completely arbitrary, we conclude that in two-dimensional flat spaces there exist infinity independent conformal transformations. In order words, there exist an infinity number of independent CKVs. Indeed, using the coordinates $(z,w)$, the metric of the two-dimensional space and its inverse are given by
\begin{equation*}
  \eta_{\alpha\beta} = \left(
                         \begin{array}{cc}
                           0 &1 \\
                           1 & 0 \\
                         \end{array}
                       \right)
                       \quad \textrm{and} \quad
                       \eta^{\alpha\beta} = \left(
                         \begin{array}{cc}
                           0 &1 \\
                           1 & 0 \\
                         \end{array} \right) \,.
\end{equation*}
Thus, in two dimensions, the constraint (\ref{Bgen}) implies that
\begin{equation*}
  \eta^{\alpha \rho}B_{\alpha \rho \beta_4\cdots \beta_p} = 0 \quad \Rightarrow \quad B_{zw\beta_4\cdots \beta_p} =0,
\end{equation*}
where the symmetry (\ref{BSym}) has been used. Taking into account the symmetry  (\ref{BSym}) and noting that in the above equation the indices $\beta_4\cdots \beta_p$ are arbitrary, we conclude that the only components of $B_{\beta_2\beta_3\beta_4\cdots \beta_p}$ that can be nonzero are those in which all indices are $z$, $B_{zz\cdots z}$, or those in which all indices are $w$, $B_{ww\cdots w}$. Then, in order to make the notation more compact, it is useful to define
\begin{equation}\label{HGB}
    H_{(q)}\equiv 2 B\hspace{-0.15cm}\underbrace{_{zz\cdots z}}_{\textrm{q times}}  \quad \textrm{and} \quad
 G_{(q)}\equiv 2 B\hspace{-0.10cm}\underbrace{_{ww\cdots w}}_{\textrm{q times}}  \,.
\end{equation}
Using this notation along with Eq. (\ref{AAgen}), we can, after some algebra, arrive at the following result:
\begin{equation}\label{AX1}
  A^{\alpha,}_{\ph{\alpha,}\beta_1\beta_2 \beta_3\cdots \beta_p} x^{\beta_1}x^{\beta_2} \cdots x^{\beta_p} =
   \delta^{\alpha}_{z}\, H_{(p-1)}\,z^{p} +   \delta^{\alpha}_{w}\, G_{(p-1)}\,w^{p} \,.
\end{equation}
Concerning the coefficient $A_{\alpha,\beta_1}$, Eq. (\ref{A2}) yields
\begin{equation}\label{AX2}
  A^{\alpha,}_{\ph{\alpha,}\beta_1} x^{\beta_1} =  (C^{\alpha}_{\ph{\alpha}\beta_1} + B \delta^\alpha_{\beta_1})x^{\beta_1} =
   \delta^\alpha_{z}( B - C_{zw} )\,z + \delta^\alpha_{w}( B + C_{zw} )\,w  \,,
\end{equation}
where it has been used the fact that $C_{\alpha\beta}$ is skew-symmetric, so that just the component $C_{zw}$ is non-vanishing. Thus, inserting the results (\ref{AX1}) and (\ref{AX2}) into the Taylor expansion (\ref{XiTaylor}), we are led to the following form for the CKV:
\begin{align}
 \bl{\xi} &= \left[ A_w +  ( B - C_{zw} )z + \frac{1}{2} H_{(1)}\,z^{2} + \frac{1}{3} H_{(2)}\,z^{3}  +  \frac{1}{4} H_{(3)}\,z^{4} +  \cdots \right]\,\partial_z \nonumber \\
 &\;\; +   \left[ A_z +  ( B + C_{zw} )w + \frac{1}{2} G_{(1)}\,w^{2} + \frac{1}{3} G_{(2)}\,w^{3}  + \frac{1}{4} G_{(3)}\,w^{4} +  \cdots \,. \right]\,\partial_w \label{Xi2D}
\end{align}
Using Eqs. (\ref{fTaylor}) and (\ref{HGB}), we obtain that the function $f$ appearing in the CKV equation is given by:
\begin{equation*}
  f = \frac{1}{2}\left[ B + H_{(1)} z + H_{(2)} z^2 + H_{(3)} z^3 + \cdots  \right] +
  \frac{1}{2}\left[ B + G_{(1)} w + G_{(2)} w^2 + G_{(3)} w^3 + \cdots  \right]\,.
\end{equation*}
Since the coefficients multiplying each power are of $z$ in Eq. (\ref{Xi2D}) are all independent from each other and arbitrary, it turns out that the function multiplying the coordinate vector field $\partial_z$ is an arbitrary function that is regular at $z=0$. Analogously, the function  multiplying the coordinate vector field $\partial_w$ in Eq. (\ref{Xi2D}) is an arbitrary function of $w$ that is regular at $w=0$. Thus, the most general CKV can be written as
\begin{equation}\label{Xifinal2D}
  \bl{\xi} = H(z) \,\partial_z \,+ \,  G(w) \,\partial_w \,,
\end{equation}
where $H(z)$ and $G(w)$ are arbitrary regular functions of their argument. The function $f$ associated to this CKV is, then, given by
\begin{equation*}
  f = \frac{1}{2}(H' + G')\,.
\end{equation*}
Due to the arbitrariness of $H$ and $G$, we conclude that the number of independent CKVs in flat two-dimensional spaces is infinity. Hence, concerning conformal symmetries and CKVs, the dimension two is quite special and do not follow that same patterns of the other dimensions.

Note that in the Euclidean signature the coordinates $z$ and $w$ defined in Eq. (\ref{zw}) are necessarily complex. This does not mean that the latter conclusions for two dimensions are not valid in Euclidean signature. Indeed, although the CKV shown in Eq. (\ref{Xifinal2D}) will be complex in such a case, we can take its real and imaginary parts, which will both be CKVs, since a linear combination of CKVs with constant coefficients is also a CKV. Thus, for any signature, the number of independent CKVs in flat two-dimensional spaces is infinity.

Since CKVs are invariant under conformal transformations, we can extend the results of the present section to conformally flat spaces, namely spaces whose metrics can be brought to a flat metric by a conformal transformation. In particular, we can state that in conformally flat spaces of dimension $n\geq3$ the number of independent CKVs is $\frac{1}{2}(n+1)(n+2)$. As we shall prove in the next section, this is the maximal number of independent CKVs in a curved space of dimension $n\geq3$. Thus, we can say that conformally flat spaces have the maximal number of independent CKVs. Then, a natural question to ask is whether the converse is true. Are there spaces with the maximal number of independent CKVs that are not conformally flat? This question will be answered in the next section. Concerning the two-dimensional case, the answer is immediate. Since every space of dimension two is conformally flat, it turns out that every two-dimensional space admit infinitely many independent CKVs.

%\begin{equation}
%  A^{\alpha,}_{\ph{\alpha,}\beta_1\beta_2 \beta_3\cdots \beta_p} x^{\beta_1}x^{\beta_2} \cdots x^{\beta_p} =  2 x^\alpha \left[ B_{zz\cdots z} (z)^{p-1} +  %B_{ww\cdots w} (w)^{p-1}\right]  -  2 z w \delta^w B^w_{\ph{w} z\cdots z} (z)^{p-2} \,.
%\end{equation}
%\begin{equation}
%   A_{\alpha,\sigma\beta_3 \beta_2 \beta_4\cdots \beta_p}  =   \eta_{\alpha\sigma} B_{\beta_3 \beta_2  \beta_4\cdots \beta_p} +  \eta_{\alpha\beta_3} %B_{\sigma \beta_2  \beta_4 \cdots \beta_p} -  \eta_{\sigma\beta_3} B_{\alpha \beta_2 \beta_4 \cdots \beta_p} \,.
%\end{equation}
%%%%%%%%%%%%%%%%%%%%%%%%%%%%%%%%%%%%%%%%%%%%%%%%%%%%%%%%%%%%%%%%%%%%%%%%%%%%%%%%%%%%%%%%%%%%%%%%
%%%%%%%%%%%%%%%%%%%%%%%%%%%%%%%%%%%%%%%%%%%%%%%%%%%%%%%%%%%%%%%%%%%%%%%%%%%%%%%%%%%%%%%%%%%%%%%%
%%%%%%%%%%%%%%%%%%%%%%%%%%%%%%%%%%%%%%%%%%%%%%%%%%%%%%%%%%%%%%%%%%%%%%%%%%%%%%%%%%%%%%%%%%%%%%%%
%%%%%%%%%%%%%%%%%%%%%%%%%%%%%%%%%%%%%%%%%%%%%%%%%%%%%%%%%%%%%%%%%%%%%%%%%%%%%%%%%%%%%%%%%%%%%%%%
%%%%%%%%%%%%%%%%%%%%%%%%%%%%%%%%%%%%%%%%%%%%%%%%%%%%%%%%%%%%%%%%%%%%%%%%%%%%%%%%%%%%%%%%%%%%%%%%
%%%%%%%%%%%%%%%%%%%%%%%%%%%%%%%%%%%%%%%%%%%%%%%%%%%%%%%%%%%%%%%%%%%%%%%%%%%%%%%%%%%%%%%%%%%%%%%%
%%%%%%%%%%%%%%%%%%%%%%%%%%%%%%%%%%%%%%%%%%%%%%%%%%%%%%%%%%%%%%%%%%%%%%%%%%%%%%%%%%%%%%%%%%%%%%%%
%%%%%%%%%%%%%%%%%%%%%%%%%%%%%%%%%%%%%%%%%%%%%%%%%%%%%%%%%%%%%%%%%%%%%%%%%%%%%%%%%%%%%%%%%%%%%%%%
%%%%%%%%%%%%%%%%%%%%%%%%%%%%%%%%%%%%%%%%%%%%%%%%%%%%%%%%%%%%%%%%%%%%%%%%%%%%%%%%%%%%%%%%%%%%%%%%
%%%%%%%%%%%%%%%%%%%%%%%%%%%%%%%%%%%%%%%%%%%%%%%%%%%%%%%%%%%%%%%%%%%%%%%%%%%%%%%%%%%%%%%%%%%%%%%%

\section{Curved Spaces with the Maximal Number of CKVs}\label{Sec.CurvedSpaces}

If $\bl{\xi}$ is a CKV then it follows that the symmetric part of its first derivative must be proportional to the metric, while there is no restriction over the skew-symmetric part. Thus, the CKV equation can be written as
\begin{equation}\label{Dxi}
  \nabla_a\xi_b = \Omega_{ab} + f g_{ab}\,,
\end{equation}
where $\Omega_{ab} = \Omega_{[ab]}$ is some skew-symmetric tensor and $f$ is some scalar function. While $f$ is related to the divergence of the vector field $\bl{\xi}$, $f = \frac{1}{n}  \nabla^a\xi_a$, $\Omega_{ab}$ is, essentially, the exterior derivative of the 1-form with components $\xi_a$. Moreover, the CKV $\bl{\xi}$ is orthogonal to a family of hypersurfaces if and only if $\Omega_{[ab}\xi_{c]}=0$.  At this point, it is also useful to introduce a notation for the gradient of $f$,
\begin{equation}\label{DF}
  F_a \equiv \nabla_a f \,.
\end{equation}
As we shall prove in the sequel, a CKV is uniquely determined once we know the values of $\xi_a$, $\Omega_{ab}$, $f$ and $F_a$ at a single point of the space.

Denoting by $R_{abcd}$ the Riemann tensor of the space, it follows that the Ricci identity reads
$$ 2\nabla_{[a}\nabla_{b]} T_{c_1c_2\cdots c_p} = R^{e}_{\ph{e}c_1ba }T_{e c_2\cdots c_p}  +
 R^{e}_{\ph{e}c_2ba }T_{c_1 e c_3\cdots c_p}  + \,\cdots\,  +  R^{e}_{\ph{e}c_p ba } T_{c_1c_2\cdots c_{p-1}e}\,, $$
where $\bl{T}$ is an arbitrary tensor. In particular, applying this identity for an arbitrary vector field $\bl{V}$, it follows from the first Bianchi identity that $\nabla_{[a}\nabla_{b} V_{c]} = 0$ holds. Applying the latter identity for the case in which the vector field is a CKV, we easily find
\begin{equation}\label{BianchiOmega}
 \nabla_{[a} \Omega_{bc]} = 0 \,.
\end{equation}
On the other hand, due to \eqref{Dxi}, the Ricci identity yields
\begin{align*}
 R^{e}_{\ph{e}cba }\xi_e &=  \nabla_a\nabla_b\xi_c - \nabla_b\nabla_a\xi_c \\
 &= \nabla_a (\Omega_{bc} + f g_{bc}) -   \nabla_b (\Omega_{ac} + f g_{ac})\\
 &= (\nabla_a \Omega_{bc} + \nabla_b \Omega_{ca}) + F_a \,g_{bc} - F_b \,g_{ac}\\
 &= (3\nabla_{[a} \Omega_{bc]} - \nabla_c \Omega_{ab}) + F_a \,g_{bc} - F_b \,g_{ac}
\end{align*}
Then, using Eq. (\ref{BianchiOmega}), we find that
\begin{equation}\label{DOmega}
  \nabla_c \Omega_{ab} = R^{e}_{\ph{e}cab }\xi_e + F_a \,g_{bc} - F_b \,g_{ac} \,.
\end{equation}
Analogously, working out the Ricci identity for the tensor $\Omega_{ab}$, we find
\begin{align}
 R^{e}_{\ph{e}cba }\Omega_{ed} +  R^{e}_{\ph{e}dba }\Omega_{ce} &=  \nabla_a\nabla_b\Omega_{cd} - \nabla_b\nabla_a\Omega_{cd} \nonumber\\
 &= \nabla_a (R^{e}_{\ph{e}bcd }\xi_e + F_c \,g_{db} - F_d \,g_{cb}) -   \nabla_b (R^{e}_{\ph{e}acd }\xi_e + F_c \,g_{da} - F_d \,g_{ca}) \nonumber\\
 &= 2 \xi_e \nabla_{[a} R^{e}_{\ph{e}b]cd}   + 2 f R_{abcd} + 2\Omega^e_{\ph{e}[a}R_{b]ecd}
 + 2g_{c[a} \nabla_{b]}F_d  -2 g_{d[a} \nabla_{b]}F_c \,, \label{DFa1}
   \end{align}
where on the second equality it has been used Eq. \eqref{DOmega}, while on the third equality Eq. \eqref{Dxi} has been used. Then, contracting \eqref{DFa1} with $g^{ad}$, we are led to
\begin{equation}\label{DFa2}
  (n-2)  \nabla_bF_c + g_{bc} \nabla^a F_a = \xi_e (\nabla^a R^{e}_{\ph{e}bca}  - \nabla_b R^{e}_{\ph{e}c}) - 2 f R_{bc}
   + \Omega^e_{\ph{e}b}R_{ec} + \Omega^e_{\ph{e}c}R_{eb} \,,
\end{equation}
where $R_{ab} = R^e_{\ph{e}aeb}$ stands for the Ricci tensor. Now, contracting the latter equation  with $g^{bc}$, we find that
\begin{equation}\label{DFa3}
  \nabla^a F_a = - \, \frac{1}{n-1} \left( f \, R  + \xi^e \nabla^a R_{ea} \right)\,,
\end{equation}
with $R = R^{a}_{\ph{a}a}$ denoting the Ricci scalar. Finally, inserting \eqref{DFa3} into Eq. \eqref{DFa2}, we eventually arrive at the following identity
\begin{equation}\label{DFa}
  \nabla_bF_c = \frac{1}{n-2}\left[  \xi_e (\nabla^a R^{e}_{\ph{e}bca}  - \nabla_b R^{e}_{\ph{e}c}) - 2 f R_{bc}
   + 2 \Omega^e_{\ph{e}(b}R_{c)e} +  \frac{1}{n-1} g_{bc} \left( f \, R  + \xi^e \nabla^a R_{ea}  \right) \right]
\end{equation}
Note that for the special case $n=2$ we cannot write $\nabla_bF_c$ in terms of $\xi_a$, $\Omega_{ab}$ and $f$, see Eq. (\ref{DFa2}), rather, just the divergence $\nabla^a F_a$ is fixed. Indeed, as we have seen in the previous section, the bidimensional case is quite special inasmuch as it allows an infinite number of independent CKVs. Thus, in the remainder of this section we restrain ourselves to the case $n\geq 3$.

Now, it is time to step aside in order to sum up what we have attained. Denoting by $\bl{\Phi}$ the list of tensor fields $\{\xi_a, \Omega_{ab}, f, F_a\}$, it follows that $\bl{\Phi}$ can have up to
\begin{equation*}
  n + \frac{1}{2}n(n-1) + 1 + n = \frac{1}{2}(n+1)(n+2)
\end{equation*}
independent components, where it has been used that $\Omega_{ab}$ is, by definition, skew-symmetric and, therefore, has $\frac{1}{2}n(n-1)$ components. Moreover, note that Eqs. (\ref{Dxi}), (\ref{DF}), (\ref{DOmega}) and (\ref{DFa}) allow us to conclude that the following relation holds
\begin{equation}\label{DPhi}
  \nabla \bl{\Phi} \sim \textrm{ Function of } \{\bl{R},   \nabla\bl{R} , \bl{\Phi} \} \,,
\end{equation}
where it has been used a schematic notation, with indices omitted and $\bl{R}$ denoting the curvature tensor and its contractions. The important point is that the first derivative of $\bl{\Phi}$ is written in terms of $\bl{\Phi}$ itself and the curvature, which is independent of $\bl{\Phi}$. Moreover, taking the derivative of the latter equation we find
\begin{equation}\label{DPhi2}
  \nabla \nabla \bl{\Phi} \sim \textrm{ Function of } \{\bl{R},  \nabla\bl{R}, \nabla\nabla \bl{R}, \nabla\bl{\Phi} \}
  \sim \textrm{ Function of } \{\bl{R},  \nabla\bl{R}, \nabla\nabla \bl{R}, \bl{\Phi} \} \,,
\end{equation}
where Eq. \eqref{DPhi} has been used in the last step. Taking successive derivatives of the latter equation and always using Eq. \eqref{DPhi} to substitute the derivative of $\bl{\Phi}$, we find the general structure
\begin{equation}\label{DPhi3}
 \nabla^{(n)} \bl{\Phi}   \sim \textrm{ Function of } \{\bl{R},  \nabla\bl{R}, \nabla^{(2)} \bl{R}, \cdots , \nabla^{(n)} \bl{R} , \bl{\Phi} \} \,,
\end{equation}
with $\nabla^{(n)}$ standing for the derivative of order $n$. Thus, once we  know the value of $\bl{\Phi}$ at a single point $p$ of the manifold, we can use Eq. \eqref{DPhi3} to find the values of all the derivatives of $\bl{\Phi}$ at $p$. Then, by means of Taylor expansion, we can obtain $\bl{\Phi}$ at all points in the neighborhood of $p$ and, eventually, obtain $\bl{\Phi}$ in the whole manifold. In particular, this means that we can obtain the components of the CKV $\xi^a$ in the whole manifold. Thus, in order to determine a CKV, we just need to know the $\frac{1}{2}(n+1)(n+2)$ components of the list $\{\xi_a, \Omega_{ab}, f, F_a\}$ at a single point. In other words, there is a one-to-one association between a CKV and the components of the field  $\bl{\Phi}$ at a point of the manifold. This proves that the maximal number of linearly independent CKVs in a manifold of dimension $n$ is $\frac{1}{2}(n+1)(n+2)$, which is the number of independent components of $\bl{\Phi}$ at a point. Note that there exists a clear correspondence between the components of $\bl{\Phi}$ and the components of the most general CKV of a flat space that we have found in the previous section. The Table \ref{Tab.Correspondence} sums up this correspondence.

\begin{table}[!htbp]
\begin{center}
\begin{tabular}{c||c|c|c|c}
  \hline
\textsc{Curved Space} & $\xi_a$  & $\Omega_{ab}$  & $f$ & $F_a$   \\ \hline
\textsc{Flat Space} & $A_\alpha$  & $C_{\alpha\beta}$ & $B$ & $B_\alpha$  \\ \hline
\textsc{Interpretation} & \small{\textrm{Translation}}  & \small{\textrm{Rotation}}  & \small{\textrm{Dilatation}} & \small{\textrm{Special Conf. Transf.}}
   \\  \hline
\end{tabular}\caption{\footnotesize{This table depicts the natural correspondences between the degrees of freedom of the most general CKV in flat space and the degrees of freedom of a CKV in a curved space. The last row shows the geometrical interpretation of the transformation generated by the CKV in flat space, which should correspond to an analogous interpretation in curved space. However, note that, differently from a flat space,  in a general curved space we cannot set specific components of the CKV to zero and isolate the action of its different parts. Indeed, a general curved space will not admit the maximal number of CKVs, so that we cannot assign arbitrary values to the components of $\bl{\Phi}$ at our will. } }\label{Tab.Correspondence}
\end{center}
\end{table}

We can say that a space admits the maximal number of CKVs when given arbitrary values for the components of $\bl{\Phi}$ at a point of the manifold there exists a CKV corresponding to this set of components. For instance, if the space admits the maximal number of CKVs then, given any point of the manifold, $p$, there exists a CKV $\bl{\xi}$ such that at this point
\begin{equation*}
  \xi_a|_p= \delta^1_a \;,\;\;  \Omega_{ab}|_p = 0 \;,\;\; f|_p = 0 \;,\;\; F_a|_p = 0 \,,
\end{equation*}
namely just the component $\xi_1$ of the list $\{\xi_a, \Omega_{ab}, f, F_a\}$ is ``turned on'' at this point. More generally, in a space with the maximal number of CKVs, any of the $\frac{1}{2}(n+1)(n+2)$ components of the list $\{\xi_a, \Omega_{ab}, f, F_a\}$ can be ``turned on'' or ``turned off'' at any point of the manifold. This fact will be of central importance to establish the integrability conditions for the existence of the maximal number o CKVs.

As a first integrability condition, note that we must have $\nabla_{[b}F_{c]} = \nabla_{[b}\nabla_{c]}f = 0$, since the Levi-Civita connection is torsionfree. Therefore, taking the skew-symmetric part of Eq. (\ref{DFa}), we find that
\begin{equation*}
  0 = \frac{1}{n-2} \left(   \nabla^a R^{e}_{\ph{e}[bc]a}  - \nabla_{[b} R^{e}_{\ph{e}c]}    \right) \, \xi_e \,.
\end{equation*}
Now, since in a space with the maximal number of CKVs the components $\xi_e$ can assume arbitrary values at any point, the latter condition implies that
\begin{equation}\label{DR1}
  \nabla^a R^{e}_{\ph{e}[bc]a}  = \nabla_{[b} R^{e}_{\ph{e}c]}
\end{equation}
must hold. This is a nontrivial integrability condition over the curvature of a space possessing the maximal number of CKVs.

Another integrability condition can be found by means of replacing the first derivatives of $F_a$ appearing in Eq. (\ref{DFa1}) by the expression (\ref{DFa}), which eventually yields the following constraint:
\begin{align}
 &2f \left\{ (n-2) R_{abcd} - 2 \left( R_{c[a}\, g_{b]d}  +  g_{c[a}\, R_{b]d} \right) +  \frac{2 R}{(n-1)} g_{c[a} \, g_{b]d} \right\}  \nonumber \\
 & + 2\xi^e \left\{(n-2)\nabla_{[a} R_{b]ecd}  - g_{c[a} \nabla^j R_{b]edj} -  g_{c[a} \nabla_{b]} R_{de}
   +  g_{d[a} \nabla^j R_{b]ecj} +  g_{d[a} \nabla_{b]} R_{ce} + \frac{2}{n-1}  g_{c[a}g_{b]d} \nabla^j R_{ej} \right\} \nonumber \\
 &  +  \Omega_{ij} \left\{ \, \delta^{[i}_a \delta^{j]}_e \left[ (n-2)  R^e_{\ph{e}bcd} + 2 g_{b[c} R^e_{\ph{e}d]} \right]
  - \delta^{[i}_b \delta^{j]}_e\left[ (n-2) R^e_{\ph{e}acd} + 2 g_{a[c} R^e_{\ph{e}d]} \right] \right. \nonumber \\
 & \quad\quad\quad\;\; + \left. \delta^{[i}_c \delta^{j]}_e\left[ (n-2) R^e_{\ph{e}dab} + 2 g_{d[a} R^e_{\ph{e}b]} \right]
  - \delta^{[i}_d \delta^{j]}_e\left[ (n-2) R^e_{\ph{e}cab} + 2 g_{c[a} R^e_{\ph{e}b]} \right]  \right\} = 0 \,. \label{IntegCondP}
\end{align}
Since in a space with the maximal number of CKVs the components of the field $\bl{\Phi}$ can take arbitrary values at any point of the manifold, it follows that each of the expressions inside the three curly brackets in the left hand side of Eq. (\ref{IntegCondP}) must be equal to zero. In particular, the vanishing of the tensor multiplied by $f$ yields
\begin{equation}\label{WeylZero}
   R_{abcd} = \frac{2}{(n-2)} \left( R_{c[a}\, g_{b]d}  +  g_{c[a}\, R_{b]d} \right) -  \frac{2 R}{(n-1)(n-2)} g_{c[a} \, g_{b]d}\,.
\end{equation}
This means that the traceless part of the Riemann tensor vanishes, namely the Weyl tensor is zero. Then, using condition (\ref{WeylZero}) into Eq. (\ref{DR1}), we end up with the following identity
\begin{equation}\label{CottonZero1}
  2\nabla_{[b} R^{e}_{\ph{e}c]} + \frac{1}{(n-1)}\,\delta^e_{\;[b}\nabla_{c]}R = 0 \,.
\end{equation}
The tensor in the left hand side of Eq. (\ref{CottonZero1}) is known as the Cotton tensor. The same conclusion could be attained by inserting the fact that the Weyl tensor vanishes into the expression being contracted with $\xi_e$ in Eq. (\ref{IntegCondP}) and equating this to zero. Analogously, the vanishing of the expression being contracted with $\Omega_{ij}$ in Eq. (\ref{IntegCondP}) can be used to prove that the Weyl tensor vanishes. Thus, summing up, we have proved that spaces with the maximal number of CKVs must have vanishing Weyl tensor and vanishing Cotton tensor.

There are several other ways to work out the integrability conditions for the existence of the maximal number of CKVs. Nevertheless, all these ways just reinforce the two integrability conditions that we have just found. Namely, the Weyl tensor and the Cotton tensor must vanish. For instance, we could workout two expressions for $2\nabla_{[e}\nabla_{d]} \nabla_c \Omega_{ab}$ and then equate both expressions. At one hand, the Ricci identity yields
\begin{align}
  (\nabla_e\nabla_d -  \nabla_d\nabla_e)\nabla_c \Omega_{ab} &=
  R^{j}_{\ph{j}cde}\, \nabla_j \Omega_{ab} + R^{j}_{\ph{j}ade}\, \nabla_c \Omega_{jb} + R^{j}_{\ph{j}bde}\, \nabla_c \Omega_{aj} \nonumber\\
 % &= R^{j}_{\ph{j}cde}\, \left( R^{i}_{\ph{i}jab }\xi_i + F_a \,g_{bj} - F_b \,g_{aj} \right) +
 % R^{j}_{\ph{j}ade}\, \left( R^{i}_{\ph{i}cjb }\xi_i + F_j \,g_{bc} - F_b \,g_{jc} \right)  \nonumber\\
 %  & \quad\quad\quad\quad\quad\quad  + R^{j}_{\ph{j}bde}\, \left( R^{i}_{\ph{i}caj }\xi_i + F_a \,g_{jc} - F_j \,g_{ac} \right)\,,  \nonumber\\
  &=   \,\xi_i  \left( R^{j}_{\ph{j}cde} R^{i}_{\ph{i}jab }  +  R^{j}_{\ph{j}bde}\, R^{i}_{\ph{i}caj } -  R^{j}_{\ph{j}ade} R^{i}_{\ph{i}cbj } \right) +
      F_j\left( g_{cb} R^{j}_{\ph{j}ade}  - g_{ca} R^{j}_{\ph{j}bde}    \right)\,,   \label{RicciIdDOmega}
\end{align}
where Eq. (\ref{DOmega}) has been used. On the other hand, one can attain another expression for the tensor in the left hand side of Eq. (\ref{RicciIdDOmega}) by means of differentiating \eqref{DOmega} twice and using Eqs. (\ref{Dxi}), (\ref{DF}),  (\ref{DOmega}) and (\ref{DFa}) after each differentiation. Equating these two expressions for $2\nabla_{[e}\nabla_{d]} \nabla_c \Omega_{ab}$ we eventually obtain the desired integrability condition. Since the final equation is terribly messy, we shall omit its explicit form here. However, what we obtain at the end is that a linear combination of terms containing the curvature tensor (and its derivatives) multiplied  by the fields $\{\xi_a, \Omega_{ab}, f, F_a\}$  must vanish. In particular, assuming that $\xi_a$, $f$ and $\Omega_{ab}$ vanish at some arbitrary point we eventually find that the Weyl tensor must vanish at his point. On the other hand, assuming that $\xi_a$, $F_a$ and $\Omega_{ab}$ are vanishing we obtain that the Cotton tensor must be zero. Thus, the imposition that the Weyl tensor and the Cotton tensor vanish form a complete set of integrability conditions, as will be clear below. Actually, for $n\geq4$ the Cotton tensor vanishes whenever the Weyl tensor is identically zero \cite{EisenhartBook}, so that, in such a case, the only integrability condition necessary for the existence of the maximal number of CKVs is the vanishing of the Weyl tensor. On the other hand, in three dimensions the Weyl tensor is zero for any space, while the Cotton tensor can be nonvanishing. Thus, in the case $n=3$, the necessary condition for the existence of the maximal number of CKVs is the vanishing of the Cotton tensor.

Now, let us interpret the results that we have just obtained. As we have seen in the previous section, $n$-dimensional flat spaces admit $\frac{1}{2}(n+1)(n+2)$ independent CKVs, thus, the maximal number of CKVs. Since the CKV equation is invariant under conformal transformations, it follows that any space that is conformally related to the flat space, namely any conformally flat space, must admit the maximal number of CKVs as well. This is quite trivial. The relevant question is whether the converse is true. Namely, must spaces with the maximal number of CKVs be conformally flat? Our calculations above reveal that the answer is yes. Indeed, in dimensions greater than four a space is conformally flat if, and only if, its Weyl tensor vanishes \cite{EisenhartBook}. In three dimensions the Weyl tensor is always zero, but not all spaces are conformally flat. A three-dimensional space is conformally flat if, and only if, its Cotton tensor vanishes \cite{EisenhartBook}. Since we have proved that spaces with the maximal number of CKVs must have vanishing Weyl tensor and vanishing Cotton tensor, \emph{we conclude that conformal flatness is a necessary and sufficient condition for a space to admit the maximal number of independent CKVs.} As acknowledged in Sec. \ref{Sec.Introduc}, this result has already been obtained in an appendix of the book \cite{EisenhartBook}. Nevertheless, the proof presented here is different and richer in details, which, hopefully, will make the result more accessible.

In most parts of this section we have ignored the case $n=2$, since this case has unique features. In particular, recall that Eq. (\ref{DFa}) is problematic in two dimensions. Nevertheless, it is interesting noticing that the conclusion of the latter paragraph is also valid for two-dimensional spaces. Indeed, in two dimensions the maximal number of independent CKVs is infinity, since this is the number of independent CKVs in flat spaces of dimension two. However, it turns out that in two dimensions every space is conformally flat \cite{Eisenhart2D}, so that, due to the conformal invariance of the CKV equation, any two-dimensional space admits the same number of CKVs as the flat space, namely infinity.

\section{Conclusions}

In this work we have presented a new proof of the following theorem: \emph{a space admits the maximal number of CKVs if, and only if, it is conformally flat}. In order to do so, we had to identify the degrees of freedom of a general CKV. Then, we took the chance to interpret the geometric role these degrees of freedom by means of comparing with the flat space case. Since the CKVs are invariant under conformal transformations, it turns out that once we know the CKVs of the flat space, we know the CKVs of any conformally flat space. Therefore, considering that in Sec. \ref{Sec.FlatSpaces} we have obtained the explicit forms of the CKVs in flat space, we can say that we have obtained the explicit forms of CKVs in all conformally flat spaces and, thus, in all spaces with the maximal number of CKVs.

\section*{Acknowledgments}
I would like to thank Conselho Nacional de Desenvolvimento Cient\'{\i}fico e Tecnol\'ogico (CNPq) for the partial financial support and to Universidade Federal de Pernambuco for the support through the Qualis A reward.

\end{document}